\documentclass[superscriptaddress,secnumarabic,amssymb,amsmath,nobibnotes,showkeys,showpacs,aps,prd,nofootinbib,twocolumn]{revtex4}
\usepackage{color}
\usepackage{graphicx}
\usepackage{dcolumn}
\usepackage{bm}

\newcommand{\beq}{\begin{equation}}
\newcommand{\eeq}{\end{equation}}
\newcommand{\bey}{\begin{eqnarray}}
\newcommand{\eey}{\end{eqnarray}}

\begin{document}

\title{Subatomic particle as the quantum universe}

\author{S S  De}
\email{ desatya06@gmail.com} \affiliation{Department of
Applied Mathematics, University of Calcutta, Kolkata 700009,
India}

\author{Farook Rahaman}
\email{rahaman@associates.iucaa.in} \affiliation{Department of
Mathematics, Jadavpur University, Kolkata 700032, West Bengal,
India}

\author{Antara Mapdar}
\email{antaramapdar@gmail.com} \affiliation{Department of
Mathematics, Shri Shikshayatan College,
11, Lord Sinha Road,  Kolkata 700071, West Bengal,
India}

\date{\today}

\begin{abstract}
Wheeler-Dewitt equation for the wave function of the universe appeared long ago in an approach to quantization of gravity. Presently, the universe considered as the microuniverse, very tiny universe compared to the outer universe to which it belongs. This tiny universe is, thus, supposed to be the subatomic particle (elementary particle) in the universe we live in. It is possible here to convert the wave function of this microuniverse satisfying Wheeler-Dewitt equation into the wave function of the quantum (subatomic) particle, which is shown to satisfy Schr\"{o}dinger equation. The inverse of the length scale indicating smallness of the particle appears in the Schr\"{o}dinger equation being the mass of the particle. Thus, the subatomic particles are shown here to be the quantum universes.

\end{abstract}


\maketitle

\section{Introduction}

                                               In quantum theoretical approach on universe, the quantum dynamics depends on Hamiltonian H and, quantization of the constraint equation on the Hamiltonian leads to the Wheeler-Dewitt (WDW) equation for the quantum state $\psi$,  the wave function of the universe. After its inception in 1967 in the article by Bryce Dewitt [1], renewed attention was started in 1983 with the important article [2] by Hartle and Hawking. Since then, it remains as an alternative approach for quantum gravity. Although, it is well known that this quantum gravitational approach through WDW equation suffers shortcomings such as the 'problem of time'; consequent attempts have been made to resolve these issues [see for example [3] and references theorem]. We shall not attend these issues here. On the other hand, we shall regard the physical variable appearing in the quantization of constraint equation as the time variable explicitly in the wave function of the universe. In fact, it will presently be considered the possibility of the quantum state of universe generated by the quantum gravitational approach on tiny universe to be regarded as the wave function of the subatomic particle within the much bigger universe, such as, the universe we live in. For this purpose, we shall follow the time reparametrization method [4] to arrive at the constraint equation which on quantization gives rise to WDW equation for the wave function of the universe. This method is, in fact, similar to Dirac's approach [5] to obtain the Schr\"{o}dinger equation which is the WDW  equation, as the operator equation on the wave function.
                         This paper is organized as follows.In section II, we shall introduce minisuperspaces for Finslerian background spacetimes of universebased on a single degree of freedom, the scale factor. In fact, we shall consider two Finslerian structures, one of which is the Finslerian perturbation of spatially flat FLRW universe. In section III, gravitational field equations for these spacetimes have been presented and, classical Hamiltonians are constructed for these cases. In the subsequent section IV , constraint equations have obtained and the WDW equations for the wave functions of the universes have been found. In the final section V, Schr\"{o}dinger equations for the wave functions of the subatomic particles have been constructed. Finally, we make some concluding remarks in section VI. We shall use Planck units $\hbar=G=c=1$.

\section{ FINSLERIAN BACKGROUND SPACETIMES OF UNIVERSE }

 ~Before introducing Finslearian structure F for the background spacetimes of universes we make some discussion on Finsler space. In 1854, Riemann  [6] suggested that the positive $n th$ root of a $ n th$ order differential form might serve as a metric funtion. In particular, the positive 4th root of a 4th order differential form can be taken as the distance element between two neighboring points, that is,
\begin{eqnarray}
\label{1}
ds^4 = g_{\mu\nu\rho\sigma}(x)dx^\mu dx^\nu dx^\rho dx^\sigma
\end{eqnarray}
Now, for the displacement along the connecting curve $ \gamma(s)$ with tangent $ y^\mu = \frac{dx^\mu}{ds}$, one can write equation (\ref{1}) as
\begin{eqnarray}
\label{2}
F_{(4)} =g_{\mu\nu\rho\sigma}(x)y^\mu y^\nu y^\rho y^\sigma
\end{eqnarray}
where $ F_{(4)}$ is the homogeneous of fourth order in $ y^\mu$'s\\
From (\ref{2}), it follows that
\begin{eqnarray*}
F_{(4)} =G_{\mu\nu}(x,y) y^\mu y^\nu
\end{eqnarray*}
where $ G_{\mu\nu}(x,y) = g_{\mu\nu\rho\sigma}(x) y^\rho y^\sigma$ is now homogeneous function of second order in $ y^\mu$'s, and if it is written as
\begin{eqnarray*}
G_{\mu\nu}(x,y) = g_{\mu\nu}(x,y) F_{(2)}(x,y)
\end{eqnarray*}
where $  F_{(2)}(x,y)$ is homogeneous of second order and $ g_{\mu\nu}(x,y)$ is a zero order homogeneous function; we have, then,
\begin{eqnarray}
\label{3}
\frac{F_{(4)}}{F_{(2)}} \equiv F^2(x,y) = g_{\mu\nu}(x,y) y^\mu y^\nu
\end{eqnarray}
where $ F(x,y) $ is new homogeneous of first order in $ y^\mu$'s .\\
In this way we arrive at the fundamental function $ F(x,y)$ respectively Finsler structure. Its dependence on the position coordinate $ x^\mu$ and to the fibor coordinate $ y^\mu $ indicates that the geometry of Finsler space is a geometry on the tangent bundle
TM. The positional dependence of Riemannian geometry is here replaced by the  pair $ (x^\mu, y^\mu)$, known as element of support, the fiber coordinates $ y^\mu$ are interpreted as internal variable, cosmic flow line, velocity of the observer, etc.\\

The homogeneous function $g_{\mu\nu}(x,y) $ obtained as above is not, in general, the Finsler metric tensor, but it simply represents a zero order homogeneous tensor for defining F. In fact, the metric tensor of Finsler space is defined as
\begin{eqnarray}
\label{4}
 g_{\mu\nu}(x,y) = \frac{1}{2} \frac{\partial^2 F^2}{\partial y^\mu \partial y^\nu},~~ y^\mu \neq0.
\end{eqnarray}
and, we can then arrive at the relation  (\ref{3}) for the Finsler structure F having the property:
\begin{eqnarray}
F(x, \lambda y)= \lambda F(x,y)~~\forall~~ \lambda>0
\end{eqnarray}
It should be noted that the distance element $ds$ is given by
\begin{eqnarray}
d s = F(x,dx)
\end{eqnarray}
Consequently, the distance traversed on the base manifold along a direction $ Y^\mu$ is given by the integral
\begin{eqnarray}
\label{7}
I = \int F(x,y) d s
\end{eqnarray}
and geodesic equation for Finsler space can be obtained by applying the principle of least action on this integral.\\
It is given by
\begin{eqnarray}
\label{8}
\frac{d^2 x^\mu}{d s^2} + 2 G^\mu = 0
\end{eqnarray}
where $G^\mu $, which is called spray coefficient of Finsler space, is given by [\ref{7}]
\begin{eqnarray}
\label{9}
G^\mu = \frac{1}{4}  g^{\mu \nu}\bigg( \frac{\partial^2 F^2}{\partial x^\lambda \partial y^\nu}y^\lambda - \frac{\partial F^2}{\partial x^\nu}\bigg)
\end{eqnarray}
The geodesic equation (\ref{8}) ensures that the Finslerian structure  F is constant along the geodesic.\\

We now introduce two particular Finslerian structures for the background spacetime of the universe. The first one is given as
\begin{eqnarray}
F^2 = y^t y^t - R^2(t)y^r y^r- r^2 R^2(t) \overline{F}^2(\theta,\phi,y^\theta,y^\phi)
\end{eqnarray}
This Finslerian structure are introduced earlier in [\ref{8}] for consideration of Finsler gravity and in [\ref{9}] for quantum gravity.\\
Here, $ \overline{F}^2$ was proposed  to be of the form:
\begin{eqnarray}
\label{11}
\overline{F}^2= y^\theta y^\theta + f(\theta,\phi)y^\theta y^\phi
\end{eqnarray}
In fact, $ \overline{F}$ is regarded as the Finsler structure of the two dimensional Finsler space. The metric tensor for the Finsler space (\ref{11}) are given by
\begin{eqnarray}
\label{12}
\overline{g}_{ij} = diag\bigg(1,f(\theta,\phi)\bigg) ~~\mathrm{and}~~ \overline{g}^{ij} = diag\bigg(1, \frac{1}{f(\theta,\phi)}\bigg)
\end{eqnarray}
For $\overline{F}$, the geodesic spray coefficients are
\[~~~~~~~~~~~~~~~~~~~~~~~~
G^2= - \frac{1}{4}\frac{\partial f}{\partial \theta}y^\phi y^\phi~~~~~~~~~~~~~~~~~~~~~~~~~~~~~~~~~~~~~~~~~~~~~ \]\begin{eqnarray}G^3= \frac{1}{4f}\bigg(2 \frac{\partial f}{\partial \theta}y^\theta y^\phi+\frac{\partial f}{\partial \phi}y^\phi y^\phi\bigg)
~~~~\end{eqnarray}
or consideration of gravity, we require a geometrical quantity which  is the Ricci scalar.In Finsler geometry there is a geometrically invariant Ricci scalar. This is given by [\ref{7}]
\[
Ric\equiv R^{\mu}_{\nu}=\] \begin{eqnarray}  \frac{1}{F^2}\bigg(2 \frac{\partial G^\mu}{\partial x^\mu} - y^\lambda \frac{\partial^2 G^\mu}{\partial x^\lambda \partial y^\mu} + 2 G^\lambda  \frac{\partial^2 G^\mu}{\partial y^\lambda \partial y^\mu} - \frac{\partial G^\mu}{\partial y^\lambda} \frac{\partial G^\lambda}{\partial y^\mu} \bigg)
\end{eqnarray}
where
\begin{eqnarray}
 R^{\mu}_{\nu} =R^\mu_{\lambda \nu \rho} y^\lambda y^\rho/F^2
\end{eqnarray}
~~~It is to be noted that although $R^\mu_{\lambda \nu \rho} $ depends on connections, $ R^{\mu}_{\nu}$ does not. Consequently, Ricci scalar depends only on the Finsler structure F and is insensitive to connections, such as Chern connection, Cartar connection etc. In fact, in Finsler gravity the vacuum field equation is given by $ Ric=0$.\\

Now, if the function f is independent of $\phi$, that is, $ f(\theta, \phi)= f(\theta)$, we can have Ricci scalar $ \overline{Ric}$ for the Finsler structure $\overline{F}$, can be found to be
\begin{eqnarray}
\overline{Ric}= -\frac{1}{2f} \frac{d^2 f}{d\theta^2} + \frac{1}{4f^2}\bigg(\frac{d f}{d \theta}\bigg)^2
\end{eqnarray}
For the constant flag curvature, that is, for $ \overline{Ric} =\lambda$ , where $ \lambda$ is a constant, we have the following equation for specification of the function $ f(\theta)$:
\begin{eqnarray}
 -\frac{1}{2f} \frac{d^2 f}{d\theta^2} + \frac{1}{4f^2}\bigg(\frac{d f}{d \theta}\bigg)^2= \lambda
\end{eqnarray}
Of course, for more general case $\lambda$ can be taken as a function of $ \theta$. For constant $ \lambda$, one can find the Finsler structure $\overline{F}$
to be as follows:
\begin{eqnarray}
\overline{F}^2 &=& \left\{
                     \begin{array}{ll}
                       y^{\theta} y^{\theta} + A \sin^{2}(\sqrt{\lambda}\theta)y^{\phi}y^{\phi},~~\textrm{if}~~ \lambda > 0, \\
                       y^{\theta} y^{\theta} + A \theta^{2}y^{\phi}y^{\phi},~~\textrm{if}~~ \lambda = 0, \\
                       y^{\theta} y^{\theta} + A \sinh^{2}(\sqrt{\lambda}\theta)y^{\phi}y^{\phi},~~\textrm{if}~~ \lambda < 0.
                     \end{array}
                   \right.
\end{eqnarray}

The constant A may be taken as unity without any loss of generality.\\
Thus, we have the following form of $ F^2$:
\begin{eqnarray}
F^2= \alpha^2 + r^2 R^2(t)\chi(\theta)y^\phi y^\phi
\end{eqnarray}
where
\begin{eqnarray}
\chi(\theta)= \sin^2\theta-\sin^2(\sqrt{\lambda} \theta)~~~\mathrm{for}  \lambda>0
\end{eqnarray}
and $\alpha$ is a Riemannaian metric, which is given by
\begin{eqnarray}
\alpha^2= y^t y^t- R^2 y^r y^r- r^2 R^2(t)(y^\theta y^\theta + \sin^2 \theta y^\phi y^\phi)
\end{eqnarray}
For the other cases, that is, for $ \lambda =0 $ and $\lambda<0 $ we can have similar forms for $ F^2$.\\

Now, we proceed to introduce the second Finslerian structure for the background spacetime of the universe, by assuming $ g_{\mu \nu \rho \sigma}(x)$ of equation (\ref{2}) to be of the form:
\begin{eqnarray}
\label{22}
g_{\mu \nu \rho \sigma}(x)=g_{\mu \nu }(x) g_{ \rho \sigma}(x)
\end{eqnarray}
For this case we can have
\begin{eqnarray}
\label{23}
F_{(4)}=\bigg(\theta(\underline{y}^2)g_{\mu\nu}(x)y^\mu y^\nu\bigg)^2
\end{eqnarray}
where $\underline{y}^2 = g_{\mu\nu}(x)y^\mu y^\nu$ and $ \theta(x)$ is a step  function given as
\begin{eqnarray}
\theta(x) &=& 1~~ \mathrm{for} ~~ x\geq 0\nonumber\\
&=&-1~~ \mathrm{for}~~  x< 0
\end{eqnarray}
From (\ref{23}), it follows that \[ F_{(4)}=g_{\mu\nu}(x,y) F_{2}(x,y)y^\mu y^\nu =\bigg(\theta(\underline{y}^2)g_{\mu\nu}(x)y^\mu y^\nu\bigg)^2 .\]
If \[F_{2}(x,y)=  g_{\mu\nu}(x,y)y^\mu y^\nu \] it follows that
\begin{eqnarray}
F^2=\frac{F_{(4)}}{F_{(2)}}= \theta(\underline{y}^2)g_{\mu\nu}(x)y^\mu y^\nu
\end{eqnarray}
For the case \[  g_{\mu\nu}(x)= \eta_{\mu \nu}g(x^0)~  ~ with ~ ~ \eta_{\mu \nu}= diag(1,-1,-1,-1),\]
we have
\begin{eqnarray}
\label{26}
F^2= \theta(\underline{y}^2)g(x^0) \eta_{\mu \nu}y^\mu y^\nu
\end{eqnarray}
This Finslerian structure was considered in [10,11] as the space of hadronic matter extension where its important was related for generation of internal symmetry of hadrons. With a pure-time transformation given as
\begin{eqnarray}
\sqrt{g(x^0)}d x^0 = d T,~~ \mathrm{or},~~ d x^0 = \frac{d T}{R(T)}
\end{eqnarray}
with $ R(T)\equiv\sqrt{g(x^0)}=\sqrt{g(t)} ,~~~~t= x^0$ in natural unit.\\
~~Here,
\begin{eqnarray}
 y^i~ \mathrm{changes~ to}~\nu^i (i=1,2,3),~~~~ \nu^0= \sqrt{g(t)} y^0
\end{eqnarray}
$F^2$ now becomes
\begin{eqnarray}
\label{29}
F^2=\theta(\underline{\nu}^2)\bigg(\nu^0 \nu^0-R^2(\tau)(\nu^1 \nu^1 +\nu^2 \nu^2 +\nu^3 \nu^3) \bigg)\nonumber\\
\mathrm{where}~~~ \underline{\nu}^2 = \nu^0 \nu^0-R^2(\tau)(\nu^1 \nu^1 +\nu^2 \nu^2 +\nu^3 \nu^3)
\end{eqnarray}
This Finslerian structure corresponds to spatially flat FRW background metric of the universe. It should be noted that from (\ref{22}) we can also arrive at the spatially flat FRW indefinite metric instead of the present definite metric (\ref{29}).\\

For the Finslerian structure given by the fundamental function F in tensors $ g_{\mu \nu}(x,y)$ and  $ g_{\mu \nu}(x,\nu)$ respectively, are given by
\begin{eqnarray}
g_{\mu \nu}(x,y) = \theta(\underline{y}^2)g(x^0) \eta_{\mu \nu}~~and,~~~~~~~~~~~~~~~~~\nonumber\\
 \textrm   ~~~~{g}_{\mu \nu}(x,\nu) = \theta(\underline{y}^2).diag \bigg(1, -R^2(\tau), -R^2(\tau), -R^2(\tau)\bigg)
\end{eqnarray}
But these tensors are not the metric tensors of the corresponding Finsler spaces. The Finslerian metric tensors can be found by using defining formula (\ref{4}). It is an interesting fact that for the above Finsler structure (\ref{26}) and (\ref{29}), it was shown in [\ref{12}] that the spray coefficients $G^\mu$ and the Christoffel symbols of the second kind, $\gamma^i_{jk}$ as calculated from the tensor (30) are the same with the respective quantities as calculated from the corresponding Finslerian metric tensors. Consequently the gravitational field equations can be obtained from these quantities, that is, from
$G^\mu$ and   $\gamma^i_{jk}$ thus constructed.

\section{ Gravitational field equations for mini-super spaces based on Finsler spaces
}
We shall first consider minisuperspace for spatially flat FLRW spacetime of scale factor a with Finslerian perturbation as given in (10) or in (19) with $R(t) = a(t)$. This Finsler space is, in fact, a $(\alpha , \beta)$-Finsler space [8]. There the modified gravitational field equations have been found with the general energy-momentum tensor for matter distribution, given by
\begin{equation}
T^{\mu }_{\nu }=\left(p_t+\rho \right)u^{\mu } u_{\nu }-p_t g^{\mu }_{\nu }+ \left(p_r-p_t\right)\eta ^{\mu } \eta _{\nu }
\end{equation}
where $u^\mu u_\mu = - \eta ^\mu \eta _\mu = 1$ ; $p_r$ , $p_t$ being respectively the radial and transverse pressures for the anisotropic fluid. These are
\begin{eqnarray}
8\pi_F G \rho &=& \frac{3\dot{a}^2}{a^2} + \frac{\lambda -1}{r^2 a^2} \\
8\pi_F G p_r &=& -\frac{2\ddot{a}}{a} -\frac{\dot{a}^2}{a^2} - \frac{\lambda -1}{r^2 a^2} \\
8\pi_F G p_t &=& -\frac{2\ddot{a}}{a} -\frac{\dot{a}^2}{a^2}
\end{eqnarray}
[   Since  the volume of Riemannian geometry may  not be equal to that of Finsler space, so, it is anodyne to mark $ 4 \pi_F $ for assigning
 the volume of  $\overline{F}$ in the field equation.
]

As in [8] we consider the barotropic equation of state
\begin{equation}
p=\omega \rho
\end{equation}
Where the pressure $p$ is composed from the pressure components $p_r$, $p_t$ and an anisotropic presure due to the anisotropic force $r^3F_a$. With the anisotropic force

\begin{equation}
F_a = \frac{2(p_t -p_r)}{r},
\end{equation}
the pressure $P$ is given as
\begin{equation}
P = (1 +\omega)p_t -\omega p_r -m^2 \frac{r^3}{2} F_a
\end{equation}
Where m is a constant having dimension of mass. With this constant m, the dimension of the third term in R.H.S. of (37) becomes that of a pressure. \\

With the equation of state (35),we have, from equations(32), (33) and (34) the following equation for the scale factor :

\begin{equation}
\ddot{a}+\frac{1+3\omega}{2}{\dot{a}^2\over a} - \frac{m^2(\lambda - 1)}{2a} = 0
\end{equation}
Presently, we shall consider case of cosmic strings for which $\omega = - {1\over 3}$, and the above equation becomes
\begin{equation}
\ddot{a} - \frac{m^2(\lambda - 1)}{2a} =0
\end{equation}
The classical Lagrangian can be constructed so that form classical action the equation can be removed. The Lagrangean is
\begin{equation}
L = {1\over 2} \ddot{a}^2 + {1\over 2} (\lambda -1) m^2 \ln(a)
\end{equation}
The classical momentum $\pi$ is given by
\begin{equation}
\pi = \frac{\partial L}{\partial \dot{a}} = \dot{a}
\end{equation}
Consequently, the classical Hamiltonian $H$ reads
\begin{equation}
H = \pi \dot{a} - L = {1\over 2}\pi^2-{1\over 2}(\lambda -1)m^2 \ln a
\end{equation}
Now we shall consider minisuperspace for the Finsler spacetime as given in (26). This spacetime has shown above to become a spatially flat FLRW Finslerian spacetime with scale factor $R(t)$. \\
It was pointed out there that the spray coefficients $G^\mu$ and the Vhristoffel symbols $\gamma^i_{jk}$ can be calculated from the metric tensor (30) instead of the Finsler metric tensor. These are given by

\begin{equation}
G^l = {1\over 2} \gamma^l_{jk}y^jy^k
\end{equation}
where
\begin{equation}
\gamma^l_{jk} = {1\over 2}\frac{g'(x^0)}{g(x^0)}\lbrace\delta^l_j\delta^0_k + \delta^l_k\delta^0_j - \eta^{l0}\eta_{jk}\rbrace
\end{equation}
Thus we have
\begin{equation}
G^l = {1\over 4}\frac{g'(x^0)}{g(x^0)}\left\lbrace 2 y^l y^0 - \eta^{l0}y^2\right\rbrace
\end{equation}
Here,
\begin{equation}
y^2 = \eta_{jk}y^j y^k
\end{equation}
Consequently,
\begin{eqnarray}
G^0 &=& {1\over 4}\frac{g'(x^0)}{g(x^0)}\left\lbrace (y^0)^2 + (y^1)^2 + (y^2)^2 +(y^3)^2 \right\rbrace \nonumber \\
G^i &=& {1\over 2}\frac{g'(x^0)}{g(x^0)} y^i y^0 , ~~~i=1,2,3 \label{Gg}
\end{eqnarray}
The ricci scalar can be obtained from these spray coefficients (\ref{Gg}) by using (14). It is given by
\begin{equation}
F^2 Ric = {1\over 2} \sum_{i=1}^3 (y^i)^2\left\lbrace {d\over dx^0} \left({g'\over g}\right)+ \left({g'\over g}\right)^2\right\rbrace - {3\over 2} (y^0)^2{d\over dx^0} \left({g'\over g}\right)
\end{equation}

As in [7] we shall follow the notion of Ricci tensor in Finsler geometry that was first introduced by Akbar-Zadeh [13], and construct the field equation in Finsler space following Li Xin, et.al.[14]. There, Ricci tensor is constructed from the Ricci scalar, Ric, which is insensitive to connections. It is given by

\begin{equation}
Ric_{\mu\nu} = \frac{\partial^2\left({1\over 2}F^2 Ric\right)}{\partial y^\mu \partial y^\nu}
\end{equation}
Also, the scalar curvature in Finsler geometry is given as
\begin{equation}
S = g^{\mu\nu} Ric_{\mu\nu}
\end{equation}
For the present Finsler space we have these quantities as
\begin{eqnarray}
Ric_{00} &=& -{3\over 2}{d\over dx^0} \left({g'\over g}\right)\nonumber \\
Ric_{ii} &=& {1\over 2}\left\lbrace{d\over dx^0} \left({g'\over g}\right) + \left({g'\over g}\right)^2\right\rbrace ~~~~ i=1,2,3
\end{eqnarray}
and
\begin{equation}
S = -3 \frac{\theta(y^2)}{g(x^0)} \left\lbrace{d\over dx^0} \left({g'\over g}\right) + {1\over 2}\left({g'\over g}\right)^2\right\rbrace
\end{equation}
The modified Einstein tensor in Finsler spacetime
\begin{equation}
G_{\mu\nu} = Ric_{\mu\nu} - {1\over 2} g_{\mu\nu}S
\end{equation}
yields
\begin{eqnarray}
G_{00} &=& {3\over 4}  \left({g'\over g}\right)^2 \nonumber \\
G_{ii} &=& -\left\lbrace{d\over dx^0} \left({g'\over g}\right) + {1\over 4}\left({g'\over g}\right)^2\right\rbrace , ~i=1,2,3
\end{eqnarray}
By writing  $g(x^0) = R^2(x^0)$, we have
\begin{eqnarray}
G_{00} &=& 3 \left({R'\over R}\right)^2 \nonumber \\
G_{ii} &=& -\left[2 {d\over dx^0} \left({R'\over R}\right) + \left({R'\over R}\right)^2\right], ~i=1,2,3
\end{eqnarray}
With the energy momentum tensor $T_{\mu\nu} = diag(\rho , p,p,p)$ and the barotropic equation of state we can have the following equation for $R(x^0)$ :
\begin{equation}
2 {d\over dx^0} \left({R'\over R}\right) + \left({R'\over R}\right)^2 \left(3\omega +1\right) = 0
\end{equation}
With the pure-time transformation (27), this equation become
\begin{equation}
\ddot{a}+\frac{1+3\omega}{2}\frac{\dot{a}^2}{a}=0
\end{equation}

for the scale factor $a(t)$ of the spatially flat FRW background Finslerian spacetime as given in (29), where $R=a$. It is interesting to note that we can arrive ar the same equation (57) by by considering Einstein field equations for the spatially flat FRW indefinite spacetime (Riemannian) with the same energy-momentum tensor and barotropic equation of state.\\
Also, it is pointed out here that the Ricci tensor $Ric_{\mu\nu}$, as in (19), was calculated from $F^2Ric$ which is given by

\begin{equation}
F^2 Ric = F^2 R^\mu_\mu = R^\mu_{\lambda\mu\rho} y^\lambda y^\rho
\end{equation}
If $R^\mu_{\lambda\mu\rho}$'s are independent of $y^\mu$'s , then we have
\begin{eqnarray}
Ric_{\lambda \rho} &&= {1\over 2} \frac{\partial^2(F^2 Ric)}{\partial y^\lambda \partial y^\rho} = {1\over 2}\frac{\partial^2(R^\mu_{\lambda\mu\rho} y^\lambda y^\rho)}{\partial y^\lambda \partial y^\rho} \nonumber \\
&& = R^\mu_{\lambda\mu\rho} = R_{\lambda\rho}
\end{eqnarray}
Consequently, the usual Einstein field equations hold good
and these are identical with the modified Einstein field equations of Finsler spacetime . This is case for the present equations in (32)-(34). Now the classical Lagrangian for the equation (57) is
\begin{equation}
L=\frac{1}{2}a^{(1+3\omega)}\dot{a^2} \label{Lag}
\end{equation}
In fact from classical action with this Lagrangian the equation (57) can be obtained . The classical momentum is
\begin{equation}
\Pi=\frac{\partial L}{\partial \dot{a}}=a^{(1+3\omega)}\dot{a^2} \label{pi}
\end{equation}
Consequently , the classical Hamiltonian H is given by
\begin{equation}
H=\Pi\dot{a}-L=\frac{1}{2}a^{(1+3\omega)}\dot{a^2} =\frac{1}{2}a^{-(1+3\omega)}\Pi^2 \label{H}
\end{equation}
\section{Constraint equation and Wheeler-Dewitt equations}
We shall follow time-reparametrization method [4] to arrive at the constraint equation from which WDW equations can be obtained . For the background Finsler spacetimes of universe , we have the equations for the scale factor a(t) and the corresponding Lagrangians $L(\frac{da}{dt},a)$ . We now introduce a new dynamical variable T instead of t and time $\tau$ for time-reparametrization . Now $T=T(c)$ and we define a new dynamical Lagrangian $L^\prime$ as
\begin{equation}
L^\prime \equiv L^\prime(\dot{T},T,\dot{a},a)=\dot{T}L(\frac{\dot{a}}{\dot{T}},a)\label{L1}
\end{equation}
Here "." represents $\frac{d}{d\tau}$ . The action S is given by
\begin{equation}
S=\int L dt \equiv \int L dT =\int d\tau \dot{T} L=\int d\tau L^\prime \label{S}
\end{equation}
The momentum
\begin{equation}
p_T=\frac{\partial L^\prime}{\partial \dot{T}}=L(\frac{\dot{a}}{\dot{T}},a)+\dot{T}\frac{\partial L}{\partial \dot{T}} \label{momen}
\end{equation}
Now,
\[
\frac{\partial L}{\partial \dot{T}}=\frac{\partial L}{\partial a^\prime}\frac{\partial a^\prime}{\partial \dot{T}}\]
where $a^\prime=\frac{da}{dT}=\frac{\dot{a}}{\dot{T}}$ . Therefore,
\begin{equation}
\frac{\partial L}{\partial \dot{T}}\dot{T}=\dot{T}(-\frac{\dot{a}}{\dot{T^2}})\frac{\partial L}{\partial a^\prime}=-\frac{\dot{a}}{\dot{T}}\frac{\partial L}{\partial a^\prime}=-a^\prime \Pi_a \label{L3}
\end{equation}
where $\Pi_a$ is momentum conjugate to a since $L(\frac{\dot{a}}{\dot{T}},a)\equiv L(a^\prime , a)$ . Consequently , the Hamiltonian associated with $L(a^\prime , a)$ is given by
\begin{equation}
H=a^\prime \Pi_a-L(a^\prime , a) \label{H2}
\end{equation}
By using (\ref{momen}) , (\ref{L3}) and (\ref{H2}) , we have
\[
p_T=L(a^\prime , a)-a^\prime \Pi_a=-H(\Pi_a,a)\]
Thus we arrive at the constraint equation
\begin{equation}
p_T +H(\Pi_a,a)=0 \label{momen3}
\end{equation}
On quantization we can have the WDW equation as
\begin{equation}
i\frac{\partial \psi}{\partial T}+\hat{H}(\hat{\Pi_a},a)\psi=0 \label{WDW}
\end{equation}
where $\hat{H}$ is now the Hamiltonian operator and $\hat{\Pi_a}=-i\frac{\partial}{\partial a}$ is the momentum operator.
It is to be noted that in Dirac's method [5] the constraint equation is regarded as an operator equation on the wave functions to obtain the WDW equation (\ref{WDW}) . It is wellknown that the constraint $H=0$ gives rise to the WDW equation which is
\begin{equation}
\hat{H}\psi=0 \label{WDW2}
\end{equation}
But we know that the classical dynamics remains unchanged by an addition of a constant to the original Lagrangian , that is , by changing Lagrangian L to $L(a^\prime , a)+E$ where E is a constant . This leads to the classical Hamiltonian H changing to $H(\Pi_a,a)-E$ . Consequently the constraint becomes $H-E=0$ , and the WDW equation reads as
\begin{equation}
(\hat{H}-E)\psi=0 \label{WDW3}
\end{equation}
Interestingly , the constraint obtained from reparametrization leads to the Schr\"{o}dinger equation (WDW equation) (\ref{WDW}) and if we put $\psi(T,a)=\phi(a)e^{iET}$ , we have from (\ref{WDW}) the following equation for $\psi$ :
\begin{equation}
\hat{H}(\hat{\Pi},a)\psi=E \psi \nonumber
\end{equation}
which is the same as the equation (\ref{WDW3}) obtained from the usual constraint equation . Further , we note that the original $\psi$ in WDW equation (\ref{WDW3}) is independent of time , where as the wave function (of the universe) $\psi$ as obtained from the constraint due to reparametrization invariance or by Dirac method , is independent on the parameter T which is interpreted as the time . Introduction of time in the wave function may also be compared with an approach by Unruh in which WDW equation are found to be
\begin{equation}
i\frac{\partial \psi}{\partial \tau}=NH\psi \label{WDW4}
\end{equation}
where $\psi$ is a function of arbitrary $\tau$ and an unknown (unmeasarable) N . Of course , we can take $N=-1$ and $\tau$ as the coordinate time .
\section{Wave equations of Fundamental particles}
For the two cases , we have to classical Hamiltonians given in (42) and (62). On quantization , we can have the respective Hamiltonian operators as follows :
\begin{eqnarray}
\hat{H}&=&-\frac{1}{2}\frac{\partial^2}{\partial a^2}-\frac{1}{2}(\lambda-1)m^2 \ln a \label{H2}\\
and ~~~\hat{H}&=&-\frac{1}{2}a^{-(1+3\omega)}\frac{\partial^2}{\partial a^2} \label{H3}
\end{eqnarray}
Therefore , for these two cases we have WDW equations as follows :
\begin{eqnarray}
i\frac{\partial \psi}{\partial t}&=&\frac{1}{2}\frac{\partial^2 \psi}{\partial a^2}+\frac{1}{2}(\lambda-1)m^2 \ln a \label{WDW5}\\
and ~~~ i\frac{\partial \psi}{\partial t}&=&\frac{1}{2} a^{-(1+3\omega)}\frac{\partial^2 \psi}{\partial a^2} \label{WDW6}
\end{eqnarray}
For $\omega=-\frac{1}{3}$ (cosmic string) , the WDW equation (\ref{WDW6}) becomes

\begin{equation}
i\frac{\partial \psi}{\partial t}=\frac{1}{2}\frac{\partial^2 \psi}{\partial a^2}
\end{equation}
For $\lambda=1$ ,the equations $(75)$ and $(77)$ are the same.In fact this case corresponds to the FRW (Riemannian) spacetime without any Finslerian perturbation. Therefore we shall first consider the WDW equation $(77)$ of the wave function of the universe in order to convert it into the Schr\"{o}dinger equation for the wave function of a particle in the universe we live in. Here the particles(subatomic) are consider as the tiny universes within much bigger universe like ours.As in Namsrai $[15]$ and also in De $[10,11]$ ,we regard the coordinate system of the tiny universe is governed by $(H,\psi)$ as described above .On the other hand the local coordinates of the outer universe , $\hat{x}^\mu$ are related to $\tau^\mu$ as
\begin{eqnarray}
\hat{x_i}&=&x_i+\tau_i ~~~,~~~  i=1,2,3 \nonumber\\
\hat{x_0}&=&x_0+\tau_0=b x_0
\end{eqnarray}
where $x_i(i=1,2,3)$ represents the spatial coordinates of the origin of the internal coordinate system with respect to the local coordinate frame of the outer universe . Here $x_0$ is the internal universe by scaling with a parameter b which will be specified later . Now we quantize $\tau^i$ according to Namsrai $[15]$ , De $[10,11]$ and De and Rahaman $[12]$ .In fact we take
\begin{equation}
\tau_i=il\sqrt{\theta y^2\theta}\gamma_i(x) ~~~,~~~  i=1,2,3
\end{equation}
where $\gamma_i(x)^,s$ are the Dirac matrices for the internal spacetime of the tiny universe , and they are related to the flat space Dirac matrices $\gamma_i$ through the vierbein $V_\mu^\alpha(x)$ . In fact the relation is
\begin{equation}
\gamma_\mu(x)=V_\mu^\alpha(x)\gamma_\alpha
\end{equation}
The flat space Dirac matrices satisfy the following anticommutation relation :
\begin{equation}
\gamma_\alpha \gamma_\beta+\gamma_\beta \gamma_\alpha=2\eta_{\alpha\beta} ~~~,~~~ where ~~ \eta_{\alpha\beta}=diag(1,-1,-1,-1)
\end{equation}
In (79) , l represent a characteristic length . Also the vierbein $V_\mu^\alpha(x)^,s$ are given by
\begin{eqnarray}
V_i^\alpha(x)&=&\sqrt{\theta( y^2)} a(t)\delta^\alpha_i  ~~~ (i=1,2,3) \nonumber\\
V_\alpha^i(x)&=&\frac{1}{\sqrt{\theta( y^2)} a(t)}\delta_\alpha^i
\end{eqnarray}
where $\theta( y^2)$ are given in (29). Then using (79) , (80) and (82) , we have from (78) the following relations
\begin{eqnarray}
\hat{x_i}&=&x_i+i \theta( y^2) l a(t) \gamma_i~~~,~~~  i=1,2,3 \nonumber\\
\hat{x_0}&=&b x
\end{eqnarray}
Now it is easy to have the relation
\begin{equation}
\frac{\partial^2}{\partial a^2}=-l^2\sum\gamma_i \gamma_j \frac{\partial^2}{\partial \hat{x_i}\partial \hat{x_j}}=l^2\sum \frac{\partial^2}{\partial \hat{x_i^2}}~~(i=1,2,3)
\end{equation}
by using the anticommutation relation (81)
Also,
\begin{equation}
\frac{\partial}{\partial t}=b\frac{\partial}{\partial \hat{x_0}}
\end{equation}
Consequently , we have from equation (77),
\begin{equation}
i t \frac{\partial \psi}{\partial \hat{x_0}}=\frac{1}{2}l^2 \sum\frac{\partial^2 \psi}{\partial \hat{x}_i^2} ~~~ (i=1,2,3)
\end{equation}
$\psi(a,t)$ is now a function $\hat{x_i}(i=1,2,3)$ and $\hat{x_0}$ , that is
\begin{equation}
\psi(a,t)\equiv \psi(\hat{x_\mu})
\end{equation}
Since it follows from (83) that
\begin{equation}
l a =\frac{1}{\sqrt 3}\sqrt{\sum(\hat{x_i}-x_i)^2} ~~~ (i=1,2,3)
\end{equation}
Now we specify the parameter b to be equal to l and the characteristic length l=$\frac{1}{m}$ where m is the mass of the particle . Then we arrive at the Schr\"{o}dinger equation for the wave function of the particle with mass m as
\begin{eqnarray}
i \frac{\partial \psi}{\partial \hat{x_0}}=\frac{1}{2m} \nabla^2\psi \\
 where ~~~~ \nabla^2\equiv \sum \frac{\partial^2}{\partial \hat{x_i^2}} \nonumber
\end{eqnarray}
For WDW equation (75) , we quantize the internal coordinates $\tau_\mu$ as $\tau_i= i l a(t)\gamma_i$ , (i=1,2,3) and consequently the relation between the local coordinates of the outer universe and $\tau_\mu$ becomes
\begin{eqnarray}
\hat{x_i}&=&x_i+i l a(t) \gamma_i~~~,~~~  i=1,2,3 \nonumber\\
\hat{x_0}&=&l x_0
\end{eqnarray}
Proceeding as before we can arrive at the following equation
\begin{equation}
i l \frac{\partial \psi}{\partial \hat{x_0}}=\frac{1}{2}l^2\sum \frac{\partial^2 \psi}{\partial \hat{x_i^2}}+\frac{1}{2}(\lambda-1)m^2 \ln a ~~~ (i=1,2,3)
\end{equation}
Hence,
\begin{eqnarray}
\psi(a,t)&\equiv &\psi(\hat{x_\mu}) \nonumber\\
and ~~~~ a&=&\frac{1}{l\sqrt 3}\sqrt{\sum(\hat{x_i}-x_i)^2} ~~~ (i=1,2,3)
\end{eqnarray}
We take $ m=m_p=1 $ (in planck unit) and , $l=\frac{1}{\bar{m}}$ where $\bar{m}$ is mass of the particle and $m_p$ is the planck mass . Then the equation (91) becomes the Schr\"{o}dinger equation for the particle of mass $\bar{m}$ :
\begin{equation}
i \frac{\partial \psi}{\partial \hat{x_0}}=\frac{1}{2\bar{m}}\nabla^2\psi+\frac{\bar{m}}{2}(\lambda-1)\ln \{\frac{1}{l\sqrt 3}\sqrt{\sum(\hat{x_i}-x_i)^2}\} ~~~ (i=1,2,3)
\end{equation}
The second term in the R.H.S. of (??) may be interpreted as the self interaction potentil arising out from the Finslerian perturbation of the minisuperspace for spatially flat FLRW universe . In fact this term vanishes for $\lambda=1$ , which is case of minisuperspace based on Riemannian spacetime , the FLRW universe of scale factor a .\\

\section{Discussion and Conclusion }

In this work, we have investigated minisuperspces as Finslerian perturbation of FLRW quantum cosmological models as well as Finslerian spacetime corresponding to spatially flat FRW background spacetime of the universe, with a single gravitational degree of freedom, the scale factor of the universe, for both the cases. For these two cases, the classical Hamiltonians have been found and, from the constraint equation we have obtained WDW equations after quantization . We, then, regarded these quantum universes to be the subatomic particles, the quantum particles (elementary particles) in the universe we live in. Actually, the WDW equations for the wave function of the universe have been converted into the Schr\"{o}dinger equations for the wave functions of the particles, the microuniverses. The extreme smallness of the tiny universes considered as subatomic particles is, here, characterized by the lengths whose inverses actually represent the masses of the particles. Thus, the present approach can be considered as a first step towards quantum gravity and, also to the simultaneous emergence of both the gravity and quantum mechanics, that has been previously advocated in an article by De and Rahaman [16].

\section*{Acknowledgments}

FR would like to thank the authorities of the Inter-University Centre for Astronomy and Astrophysics, Pune, India for providing research facilities.   FR is also grateful to DST-SERB,  Govt. of India  and  Jadavpur University for financial support under RUSA 2.0.

\end{document}